       \documentclass{aa}     
       \usepackage{txfonts} 
       \usepackage{graphicx} 
       \usepackage{natbib} 

\begin{document} 

       %
        \title{Chemical composition of evolved stars of high galactic latitude 
      \thanks{Based on observations obtained at the Haute-Provence Observatory, France.}} 
      \markboth{Abundances}{Arellano Ferro} 

       \markboth{Abundances}{Arellano Ferro} 
          \author{Sunetra Giridhar$^1$, A. Arellano Ferro$^2$} 
          \institute{                $^1$Indian Institute of Astrophysics,  Bangalore 560034, 
       India\\giridhar@iiap.ernet.in\\ 
      $^2$Instituto de Astronom\'{\i}a, Universidad Nacional Aut\'onoma de 
       M\'exico,\\ 
                      Apdo. Postal 70-264, 
       M\'exico D.F. CP 04510; \\armando@astroscu.unam.mx \\} 

      \maketitle 

        \begin{abstract} 
      %

   We have  carried out abundance analysis for a sample of high galactic
 latitude supergiants in search of evolved stars. 
      We find that HD 27381 has atmospheric parameters and an abundance pattern 
           very similar to that of the post-AGB star HD 107369. 
            HD 10285 and HD 25291 are moderately metal-poor and 
            show the influence of mixing that has brought the products 
           of NeNa cycle to the surface. 
          The high galactic latitude B supergiant HD 137569 shows selective  
       depletion of refractory elements normally  seen in  post-AGB stars.
         We find that the high velocity B type 
        star HD 172324 shows moderate deficiency of Fe group elements but 
            the CNO abundances are very 
           similar to that of disk B supergiants. The observed variations in 
    the radial velocities, transient appearance of emission components
     in hydrogen line profiles  and doubling 
           of O~I lines at 7774\AA~ support the possibility of this star being 
           a pulsating variable or a binary star. 

        \end{abstract} 

      %

      \section{Introduction} 

         There has been considerable interest in the chemical composition 
               studies of stars in various stages of evolution, as they provide 
               diagnostic tools to examine the predictions made by  theories of 
     stellar evolution. Of particular interest among evolved stars are the 
               post-AGB stars, where the abundance peculiarities resulting from 
     evolutionary processes manifest on a relatively short time scale. 
   A fairly large number of post-AGB stars have been studied in the last
   two decades. Within them, 
 a few distinct subgroups have been identified. One group containing 
      stars with  C/O $>$ 1 displays considerable 
      enhancement of s-process elements. 
               Typical examples are HD 56126, HD 187785 and IRAS06530-0213 
              (see Reddy et al. 2002; and Van Winckel 2003 for an overview). 
  These objects show two peaks in their Spectral Energy Distribution 
               (SED) and their metallicity 
     ranges between  [Fe/H]=$-$0.3 and $-$1.0. The O-rich post-AGB stars 
    (C/O $<$  1) do not display s-process enhancement but display a 
   double humped SED  and  metallicity range similar to that seen in C-rich AGB stars. 
  The typical examples are 89 Her, HD 161796, HD 133656, SAO 239853 etc. 
   Yet another group of post-AGB stars with C/O $\sim$ 1   
    show abundance peculiarities caused by selective removal of 
        condensable elements e.g. BD$+$39$^{o}~$4926, 
              HR 4049, HD 44179, HD 46703, HD 52961 
      (references to the individual objects can be found in a review paper 
    by Van Winckel 2003).  Although a large fraction of them are known to 
    be binaries and the presence of IR fluxes lend further support to the 
  idea of circumbinary disk as the site of dust-gas separation,
 lack of detected  IR for  BD$+$39$^{o}~$4926 is difficult to explain
 via a binary hypothesis.  Many RV Tau  stars like AD Aql,
 AC Her, EP Lyr, AR Pup,  UY CMa, HP Lyr, DY Ori, BZ Sct, RU Cen and SX Cen 
 (see Giridhar et al. 
1998, 2000, 2005;  Maas et al. 2002 and references therein)   
 also show very similar abundance patterns. Although AC Her, V Vul, RU Cen 
 and SX Cen are  RV Tauri stars with a detected binary companion,
  the frequency of  binaries in  RV Tauri stars is yet to be determined. 

   A fourth group contains hot post-AGB stars which are metal-poor   
   with remarkably high carbon deficiency (McCausland et al. 1992, 
  Conlon et al. 1991,  Moehler \& Heber 1998,  Mooney et al. 2002). 
  These stars also show  considerable deficiency of N, O, Mg and Si 
   although these deficiencies are not as large as seen for carbon. 
  Examples of this group are PG 1323-086, LS IV $-4^{o}.01$ and LB 3139. 
  These objects are believed to be evolving on low mass post-AGB 
   evolutionary tracks. 

    Within post-AGB stars, those evolving from relatively massive main sequence 
    stars (  $\sim 7M_{\odot}$) would have a shorter transition time
            before becoming (Planetary Nebulae) PNe and 
           would be surrounded by dense circumstellar shells. 
             The very stringent criteria of low gravity, 
       high galactic latitude, strong IR excess and photometric variabilities 
             given in Trams et al. (1991) and their 
      location in the IRAS two color diagram of van der Veen   
    \& Habing (1988) has been extremely useful in detecting the new 
     C and O-rich post-AGB stars described above. However, there are
    interesting stars like  BD$+$39$^{o}~$4926 
   with no IR detection and other hot post-AGB stars that do not exhibit large 
    IR excess. It is likely that these stars have evolved  from the low-mass 
  end of post-AGB domain (1.5 to 3M$_{\odot}$). These stars would have 
   longer transition times and would have lost most of
 their circumstellar envelope before moving towards the PNe region.
  With a longer transition period, their detection probability
             would be higher.   
                                
        Our sample, therefore, contains high galactic latitude, low gravity 
   stars without insistence on excess IR. Although HD 10132, HD 25291 and HD 27381 do 
  have larger than stellar fluxes at 100$\mu$  wavelengths, in most cases  
  the excess IR is smaller than the measurement errors. Similarly for HD 12533
  and HD 159251 the observed SED indicates  that the observed IR fluxes do not translate into
  IR excess larger than the measurement errors.  A few high velocity and 
  high galactic latitude  supergiants like HD 10285,
    HD 137569  and HD 172324 without IR detection are also
    included but these are known to show light variabilities. 
      In a previous paper (Arellano Ferro, Giridhar \& Mathias 
     2001; Paper I) we have presented  our detailed abundance analysis
 for the interesting post-AGB stars  HD 158616 and HD 172481.
   In the same paper, we had reported metal and carbon deficiency for
   HD 172324 and suggested that a more comprehensive analysis 
   was required. With extended spectral coverage we have now made a
    more complete analysis of this object. 

      The present paper 
      is organized as follows: In section \ref{obs} the observations and 
       reduction techniques are briefly described and 
    a discussion of photometric estimates of the initial values of
 the effective temperature and the uncertainties in the derived abundances 
   is given. Section \ref {specana} contains the detailed discussion
 on the elemental abundances of each 
  object in the sample.  Section \ref{conclusions} summarizes our
 findings and derived conclusions.

      \section{Observations and data reduction}\label{obs} 

    Table 1 contains the list of stars studied in this work, their spectral 
    types, magnitudes, galactic positions and, if available, 
    the IRAS infrared fluxes. 

    The observational material for this work was obtained  during 
      October 5 - October 10, 2000  with the 1.93m telescope 
      of the Haute-Provence Observatory (OHP), 
   which is equipped with the high resolution (42000) echelle  spectrograph
  ELODIE. Details about the  performance and characteristics of the 
   instrument have  been thoroughly described by Baranne et al. (1996). 
   These spectra were reduced using spectroscopic 
    data reduction tasks available in the IRAF package.  At our request, an 
 additional spectra  of  HD 172324 and HD 137569
  were obtained  by Dr. E. Reddy and Dr David Yong using the 2d Coud\'e 
  echelle spectrograph of  2.7m telescope at McDonald observatory described in
  Tull et al (1995). 
     These spectra have resolution of 60000. 


      \begin{table*} 
      \caption{Basic data for the sample stars}                                                                 
      \begin{center} 
      \begin{tabular}{llrrrrrrrr} 
      \noalign{\smallskip}                                                               
      \hline       
      \noalign{\smallskip}                                                               
      \noalign{\smallskip} 
      \multicolumn{1}{c}{Star}& 
      \multicolumn{1}{c}{Sp.T.}& 
      \multicolumn{1}{c}{$V$}& 
      \multicolumn{1}{c}{$l$}& 
      \multicolumn{1}{c}{$b$}& 
      \multicolumn{1}{c}{IRAS}& 
      \multicolumn{1}{c}{12$\mu$}&   
      \multicolumn{1}{c}{25$\mu$}&   
      \multicolumn{1}{c}{60$\mu$}&   
      \multicolumn{1}{c}{100$\mu$}\\ 
      \noalign{\smallskip} 
      \multicolumn{1}{c}{}& 
      \multicolumn{1}{c}{}& 
      \multicolumn{1}{c}{$(mag.)$}& 
      \multicolumn{1}{c}{$(^o)$}& 
      \multicolumn{1}{c}{$(^o)$}& 
      \multicolumn{1}{c}{}& 
      \multicolumn{1}{c}{$(Jy)$}&   
      \multicolumn{1}{c}{$(Jy)$}&   
      \multicolumn{1}{c}{$(Jy)$}&   
      \multicolumn{1}{c}{$(Jy)$}\\ 
      \noalign{\smallskip}   
      \hline   
      \noalign{\smallskip}   

      HD 10132& G5 &7.75&132.57 & $-20.3$&02008+4205&.69& .39& .40& 1.00\\ 
      HD 10285& A5 &8.60&128.59& +1.12&&&&&\\ 
      HD 12533&K3IIb&2.26&136.96 & $-18.56$&01285+6115&98.55&23.96&3.59& 1.16\\ 
      HD 25291&F0II&5.04&145.51& $+5.03$&04002+5901&1.60& .41& 0.40L&2.09L\\ 
      HD 27381&F2&7.55&161.64&$-8.07$&04175+3827&.81&.32L&.40L&2.19L\\ 
      HD 137569&B5III:&7.91&21.87& $+51.93$&& &&\\ 
      HD 159251&G5&7.24&117.33&$+29.32$&17193+8439&.52& .16& .40L&3.78L\\ 
      HD 172324&B9Ib&8.16&66.18 &$+18.58$&&&&&\\ 
                 \noalign{\smallskip}   
                 \hline   
                 \noalign{\smallskip}   
      \end{tabular} 
      \end{center} 
      \end{table*}   

      \subsection{Spectroscopic reductions} 

  For details about the reductions of our spectra, from the raw CCD images 
    to the 
   measurement of equivalent widths and a discussion of their uncertainties, 
      the reader is refered to Paper I. 

      \subsection{Determination of atmospheric parameters} 

      As is well-known, the line strengths are  affected by 
      atmospheric parameters like the  effective temperature 
      ($T_{\rm eff}$), gravity (log $g$) 
      and microturbulent velocity ($\xi$). 
      It is  therefore necessary to determine these parameters before 
      using line strengths for abundance determinations. 

      The present sample contains stars with a large range in temperatures. 
      The criteria used for determining the atmospheric parameters for hot 
       members ($T_{\rm eff}$ $>$ 10000~K) were different from 
       those at the cooler side. For stars cooler than 8000~K 
       the Fe-group elements like Ti, Cr and Fe have a large number 
       of lines covering a range in line strengths, excitation potential 
       and two stages of ionization. We have therefore followed 
       the standard procedure of estimating $\xi$ by requiring that the derived 
 abundances are independent of line strengths. The temperature and gravity 
 were estimated by eliminating the dependence of computed abundances on 
 lower excitation potentials and requiring that the neutral and ionized 
 lines give the same abundances. 

  Starting values of $T_{\rm eff}$ and log $g$ are obtained from
  the photometric indices. For stars of intermediate temperatures, 
 we possess unpublished empirical calibrations of reddening-free Str\"omgren 
 photometric indices [$m_1$], [$c_1$] and H$_\beta$, 
 in terms of $T_{\rm eff}$. 
 These calibrations were calculated for 41 stars with spectral 
  types between A0 and K0 and luminosity classes I or II.
 The effective temperatures of calibrating stars  
        were determined from 13-color photometry 
      (Bravo-Alfaro, Arellano Ferro \& Schuster, 1997). 

       For hotter members we used line strengths 
       and profiles of hydrogen and helium lines and 
       the ionization equilibrium of Si~I/Si~II, Mg~I/Mg~II whenever the related 
       spectral data were available. 

       We have used the 2002 version of the spectrum synthesis code 
       MOOG written by Sneden (1973) 
       in both line and spectrum synthesis mode for stars cooler than 8000~K. 
       We used LTE model atmospheres of Kurucz (1993), and the 
      revised list of oscillator strengths of Luck (2002) 
      for Fe-group elements. For lighter  elements we used log $gf$ 
       values compiled by Wiese et al. (1996). For Fe~I lines we 
  preferred log $gf$ values given in Lambert et al. (1996) or Luck (2002) for 
  Fe~II we used the data of Giridhar \& Arellano Ferro (1995) and those in 
      Table A2 of Lambert et al. (1996). For heavier elements 
       we used the Vald-2 database (Kupka et al. 1999) 
      in addition to Luck (2002).

      \begin{table*} 
     \caption{ Derived physical parameters for program stars}                                                                 
      \begin{center} 
      \begin{tabular}{lccccc} 
      \noalign{\smallskip}                                                               
      \hline       
      \noalign{\smallskip}                                                               
      \noalign{\smallskip} 
      \multicolumn{1}{c}{Star}&     
      \multicolumn{1}{c}{$T_{\rm eff}$}&   
      \multicolumn{1}{c}{$\log g$}& 
      \multicolumn{1}{c}{$\xi$}& 
      \multicolumn{1}{c}{$V_r (hel)$ }& 
      \multicolumn{1}{c}{$V (LSR)$}\\ 
      \noalign{\smallskip} 
      \multicolumn{1}{c}{}&     
      \multicolumn{1}{c}{(K) }&   
      \multicolumn{1}{c}{}& 
      \multicolumn{1}{c}{$\rm (km~s^{-1})$}& 
      \multicolumn{1}{c}{$\rm (km~s^{-1})$ }& 
      \multicolumn{1}{c}{$\rm (km~s^{-1})$}\\ 
                 \noalign{\smallskip}   
                 \noalign{\smallskip}   
                 \hline   

      HD 10285&7750 & 1.5 & 3.5 &$-69.9$&$-64.2$\\ 
      HD 25291&7250 & 1.5 & 3.5 &$-19.2$&$-18.3$\\ 
      HD 27381&7500 & 1.0 & 4.0 &$-5.4$&$-11.4$\\ 
      HD 137569&12500 & 3.0 & 8.0 &$-45.7$&$-30.2$\\ 
      HD 137569$^\dag$&12000 & 3.0 & 7.0 &$-55.2$&$-39.6$\\ 
      HD 172324$^*$&11000& 2.5&5.0&$-126.1$ &$-106.4$\\ 
      &11500&2.5&7.5&$-117.3$&$-97.5$ \\ 
      &10500& 2.5&5.4 &$-139.0$&$-119.3$\\ 
      &10500& 2.5&6.8 &$-105.2$&$-85.7$\\ 
      && &&$-152.1$&$-132.5^+$\\ 
                 \noalign{\smallskip}   
                 \hline   
                 \noalign{\smallskip}   
      HD 10132&5000 & 2.0 & 1.0 &$-70.2$&$-64.5$\\ 
      HD 12533&4250 & 2.0 & 1.8 &$-11.2$&$-10.9$\\ 
      HD 159251&4800 & 2.75 & 1.2 &$-39.2$&$-27.6$\\ 
                 \noalign{\smallskip}   
                 \hline   
                 \noalign{\smallskip}   
      \end{tabular} 

      $\dag$ -- From McDonald spectrum obtained in 2004. 
      * -- Four independent analyses carried out for 
      different epochs are included. The first two are from Paper I. 
      The second two are from the present work. 
      +  Radial velocity measured on a 17000 resolution spectrum obtained at 
      San Pedro M\'artir Observatory on November 18, 2000. 
      \end{center} 
      \end{table*}

      \subsection{Uncertainties in the elemental abundances}\label{uncer}   

      For most of the spectra  employed in the 
      abundance analysis the S/N ratio is in the 50-100 range. 
      For the stars with 4500~K $<T_{\rm eff} <$ 8000~K we could find 
       clean unblended lines for a large number of elements and we 
       believe that the measured equivalent widths have an accuracy of 5-8\%. 
        The sensitivity of derived abundances to 
       changes in the model atmosphere parameters are described in Table 3 
       using a small set of lines with well determined atomic data. 
       The last column of Table 3  shows  the errors corresponding to an 8\%
      error in equivalent width  measurements. It is representative 
      of the  basic precision that can be attained with the instrumental setup. 
       We  find that at $T_{ \rm eff}$  of 5000~K, a change in temperature of
       250~K causes the change in line strength which is larger than
       the measurement errors of line strengths. These changes have been
      presented as changes in abundances caused by $\Delta$ $T_{\rm eff}$
       in column  6 of  Table 3. Similarly, columns 7 and 8 are indicative of
      sensitivities of line strengths to changes in log $g$ by 0.5 and $\xi$ by
       0.5~km~s$^{-1}$. The spectra of HD 10132, HD 12533 and HD 159251 had
   clean unblended lines for all representative atoms covering a good range
   in equivalent widths, excitation potential and two stages of ionizations
       for many elements (Sc, Ti, V, Cr, Fe).
       Hence for these three stars the atmospheric parameters can be measured
       with accuracies indicated in Table 3.
       
   HD 25291, HD 10285 and HD 27381 have temperatures in 7000~K to 7500~K
    range. For these stars, the spectra are 
   much cleaner, but not as scanty as to prevent us from getting the required
   range in line strengths, excitation potentials and two stages of ionisation.
   We have used neutral and ionised lines of Mg, Si, Cr and Fe to derive
   the gravities. For these stars the temperatures are not large enough for
   the lines to develop strong wings making line strengths inaccurate.
   At the temperature of 7500~K, we find that the line strengths change 
   by a larger amount if the temperature is changed by 250~K. These changes
  (presented as changes in abundances) are much larger than the measurement
    errors of the line strengths and therefore are easily discernible. 
   We are limited by the coarseness of the grid of model atmospheres. 
  A change of log $g$ by 0.5 causes changes in line strengths that are twice
  the error of line strength measurements and hence is easily discernible.
 We have used additional temperature  and gravity constraints such as the
 strengths and profiles of H$_{\gamma}$ and  H$_{\delta}$       
   to further improve the accuracy of these parameters.
   We have shown in Figure 1 the loci of temperatures and gravities 
   derived using different criteria. The adopted value of temperature and
   gravity is indicated by the asterisk.
   The gravities are estimated with an accuracy of $\pm 0.25$ dex 
    through profile fitting of H$_{\gamma}$ and H$_{\delta}$ lines.
    At 7500~K, as one can see from Table 3, the calculated abundances are 
      less sensitive to the changes in 
       microturbulent  velocities and hence our estimate of $\xi$ may have 
      an uncertainty of 1-2 km~s$^{-1}$.

      The errors in $gf$ values vary from element to element. For Fe I lines, 
      experimental values of accuracies better than 5\% do exist, 
      for other Fe-peak    elements the range in errors could be
     within  10 to 20\%. For heavier elements, 
  particularly for $s$-process elements, the errors could be larger than 25\%. 
      Hence for Fe-peak elements with a large number of lines measured 
      and good $gf$ estimates available, abundances 
      can be estimated with an accuracy of  0.15 to 0.2~dex. For the elements 
      where fewer lines are available, like s-process elements, the uncertainty 
      could be $\pm$0.3~dex or more. 

         We therefore believe that our abundance estimates for stars HD 10132, 
      HD 12533, HD 159251, HD 10285, HD 25291 and HD 27381 have accuracies
     within 0.2~ and 0.4~dex depending upon the number of lines measured and the
      quality of the $gf$ values available. 

      For stars hotter than 10000~K the lines are fewer and have
       shallower profiles with extended wings. 
      Hence the line strengths cannot be measured as accurately. 
 We have therefore estimated the errors in the calculated abundances caused by 
      the errors of equivalent measurements using 16\% as the typical accuracy 
      of line strength measurements. Since the spectrum synthesis code MOOG 
      was primarily written for late type stars and does not incorporate 
      the  opacities necessary for the flux computation of hot stars, 
      we chose to use the SPECTRUM code which is 
      described in Gray \& Corbally (1994) for the analysis of 
      hot stars HD 137569 and HD 172324. 
     We were concerned about the systematic differences that 
      might be caused by the use of these two different codes. We have therefore 
      computed the spectrum of Vega using a stellar model with $T_{\rm eff}$ 
       of 9400~K, log $g$ of 3.9 and $\xi$ of 2.0~km~s$^{-1}$ 
       and the recent abundance estimates from the literature. 
    The  spectra computed using MOOG and SPECTRUM codes 
    were compared with the high resolution spectrum 
      of Vega (Qiu et al. 1999a,b)  in selected spectral regions containing  
      important lines for many elements of interest. We found 
      very satisfactory agreement for all elements with the
   the abundances derived using the two codes up to temperature of 8000K.
    Hence we believe there is no systematic difference in estimated abundances
    caused by the use of two different codes. 
       However at temperatures hotter than 10000~K the departure from 
       LTE becomes important.  We have applied the systematic corrections 
      caused by the neglect of non-LTE and they are discussed in the subsection 
      of each star. These effects and the relative paucity of lines
    of representative atoms results in larger measurement errors for the
    abundances for HD 172324 and HD 137569. Hence their estimated
   abundances may be accurate to  only to one significant digit.

           \begin{table*} 
   \caption{Sensitivity of the calculated abundances to the changes in the 
           atmospheric parameters } 
      \scriptsize{ 

           \begin{center} 
           \begin{tabular}{llccccccc} 
  HD 10132 &Adopted & parameters &T$_{eff}$&= 5000~K& log g= 2.5& $\xi=1.0~kms^{-1}$& &\\ 
           \noalign{\smallskip}
           \hline 
           \noalign{\smallskip} 
           \noalign{\smallskip} 
           \multicolumn{1}{l}{Wavelength}& 
           \multicolumn{1}{l}{Species}& 
           \multicolumn{1}{c}{EP}& 
           \multicolumn{1}{l}{W$_\lambda$mA} & 
           \multicolumn{1}{c}{$log$ $\epsilon$}& 
           \multicolumn{1}{c}{$\Delta T_{_eff}+250~K $}& 
           \multicolumn{1}{c}{$\Delta log g +0.5$}& 
           \multicolumn{1}{c}{$\Delta  \xi +0.5$} & 
           \multicolumn{1}{c}{$\Delta  W_\lambda  +8\%$}\\ 
           \noalign{\smallskip} 
           \hline 
           \noalign{\smallskip} 
           \noalign{\smallskip} 
    5867.57 & CaI & 2.93& 43.2 & 6.55 & +0.20 & $-0.20$ & $-0.06$ &+0.08 \\   
    6169.04 & CaI & 2.52& 116.5 & 6.48 & +0.27& $-0.27$& $-0.19$&+0.13 \\ 
    5441.32& FeI& 4.31&  52.0& 7.47& +0.22& $-0.19$& $-0.12$& +0.10 \\ 
    6054.07& FeI& 4.37&  24.0& 7.58& +0.19& $-0.18$& $-0.04$& +0.05 \\   
    4631.48& FeI& 4.37&  35.3& 7.64& +0.19& $-0.19$& $-0.07$& +0.06 \\ 
    5294.56& FeI& 3.64&  32.2& 7.56& +0.23& $-0.23$& $-0.06$ & +0.06 \\ 
    5320.03& FeI& 3.64&  42.0&  7.46& +0.24& $-0.23$& $-0.09$ & +0.07 \\ 
    6411.66& FeI& 3.65& 152.0&  7.46& +0.30& $-0.28$& $-0.17$& +0.10 \\ 
   6084.12& FeII& 3.20&  42.3& 7.41& $-0.08$& +0.14& $-0.11$& +0.07 \\ 
   6149.25& FeII& 3.89&  52.6&  7.38& $-0.11$& +0.18& $-0.16$& +0.10 \\ 
    6204.64&  NiI& 4.09&  39.6&  6.17& +0.19& $-0.14$& $-0.08$& +0.07 \\   
    6175.42&  NiI& 4.09&  66.0&  6.18& +0.22& $-0.16$& $-0.19$& +0.12 \\ 
           \noalign{\smallskip} 
           \hline 
           \noalign{\smallskip} 
           \noalign{\smallskip} 
  HD 27381 &Adopted & parameters &T$_{eff}$&= 7500~K& log g= 1.0& $\xi=4.0~kms^{-1}$& &\\ 
           \noalign{\smallskip} 
           \noalign{\smallskip} 
           \hline 
           \noalign{\smallskip} 
           \noalign{\smallskip} 
           \multicolumn{1}{l}{Wavelength}& 
           \multicolumn{1}{l}{Species}& 
           \multicolumn{1}{c}{EP}& 
           \multicolumn{1}{l}{W$_\lambda$mA} & 
           \multicolumn{1}{c}{$log$ $\epsilon$}& 
           \multicolumn{1}{c}{$\Delta T_{_eff}+250~K $}& 
           \multicolumn{1}{c}{$\Delta log g +0.5$}& 
           \multicolumn{1}{c}{$\Delta  \xi +0.5$} & 
           \multicolumn{1}{c}{$\Delta  W_\lambda  +8\%$}\\ 
           \noalign{\smallskip} 
           \noalign{\smallskip} 
           \hline 
           \noalign{\smallskip} 
           \noalign{\smallskip} 
  6122.23 & CaI & 1.89& 40.2 & 5.81 & +0.44 & $-0.20$ & $-0.01$ &+0.04 \\   
  6162.18 & CaI & 1.90& 56.2 & 5.78 & +0.45 & $-0.21$ & $-0.02$ &+0.04 \\   
  6462.57 & CaI & 2.52& 44.2 & 5.78 & +0.43 & $-0.21$ & $-0.02$ &+0.05 \\   
  5328.04 & FeI & 0.93& 140.2 & 6.88 & +0.40 & $-0.14$ & $-0.09$ &+0.12 \\   
  5569.62 & FeI & 3.42&  34.0 & 6.80 & +0.30 & $-0.15$ & $-0.01$ &+0.04 \\   
  5572.84 & FeI & 3.40&  55.0 & 6.84 & +0.34 & $-0.15$ & $-0.01$ &+0.05 \\   
  5162.27 & FeI & 4.18&  44.0 & 6.96 & +0.32 & $-0.15$ & $-0.01$ &+0.05 \\   
  5415.27 & FeI & 4.39&  63.0 & 6.91 & +0.32 & $-0.15$ & $-0.02$ &+0.05 \\   
  4620.51 & FeII & 2.83& 127.0 & 6.82 & +0.16 & +0.08 & $-0.08$ &+0.09 \\   
  4632.68 & FeII & 2.89&  94.0 & 6.89 & +0.17 & +0.07 & $-0.04$ &+0.06 \\   
  6419.25 & FeII & 3.89& 105.0 & 6.89 & +0.15 & +0.08 & $-0.04$ &+0.10 \\   
   \noalign{\smallskip} 
           \noalign{\smallskip} 
           \hline 
           \noalign{\smallskip} 
           \noalign{\smallskip} 
   HD 137569 &Adopted & parameters &T$_{eff}$&= 12000~K& log g= 2.5& $\xi=7.0~kms^{-1}$& &\\ 
           \noalign{\smallskip} 
           \noalign{\smallskip} 
           \hline 
           \noalign{\smallskip} 
           \noalign{\smallskip} 
           \multicolumn{1}{l}{Wavelength}& 
           \multicolumn{1}{l}{Species}& 
           \multicolumn{1}{c}{EP}& 
           \multicolumn{1}{l}{W$_\lambda$mA} & 
           \multicolumn{1}{c}{$log$ $\epsilon$}& 
           \multicolumn{1}{c}{$\Delta T_{_eff}-500~K $}& 
           \multicolumn{1}{c}{$\Delta log g +0.5$}& 
           \multicolumn{1}{c}{$\Delta  \xi +1.0$} & 
           \multicolumn{1}{c}{$\Delta  W_\lambda  +16\%$}\\ 
           \noalign{\smallskip} 
           \hline 
           \noalign{\smallskip} 
  7423.64 & N I &10.33& 33.8 & 6.55 & +0.15 & +0.33 & $-0.01$ &+0.09 \\   
 7442.29 & N I &10.33& 50.0 & 6.49 & +0.16 & +0.33 & $-0.02$ &+0.11 \\   
  7468.31 & N I &10.33& 66.4 & 6.53 & +0.16 & +0.33 & $-0.02$ &+0.12 \\   
  5330.74 & O I &10.74& 36.0 & 6.92 & +0.14 & +0.32 & $-0.02$ &+0.10 \\   
  6155.99 & O I &10.74& 39.0 & 6.89 & +0.16 & +0.33 & $-0.02$ &+0.10 \\   
  6156.78 & O I &10.74& 62.0 & 7.00 & +0.15 & +0.32 & $-0.03$ &+0.11 \\   
  5014.03 & SII &14.07& 28.6 & 6.98 & +0.11 & +0.25 & $-0.01$ &+0.09 \\   
  5432.80 & SII &13.62& 23.0 & 6.51 & +0.13 & +0.26 & $-0.03$ &+0.08 \\   
  5606.11 & SII &13.73& 20.0 & 6.68 & +0.13 & +0.28 & $-0.03$ &+0.10 \\   
  6578.05 & CII &14.45& 36.0 & 4.10 & +0.34 & +0.69 & $-0.04$ &+0.13 \\   
  6582.88 & CII &14.45& 19.9 & 3.92 & +0.34 & +0.70 & $-0.02$ &+0.12 \\   
           \noalign{\smallskip} 
           \hline 
           \noalign{\smallskip} 
           \end{tabular} 
           \end{center} 
      } 
           \end{table*}

            \begin{figure}![h] 
            \includegraphics[width=8.cm,height=10.cm]{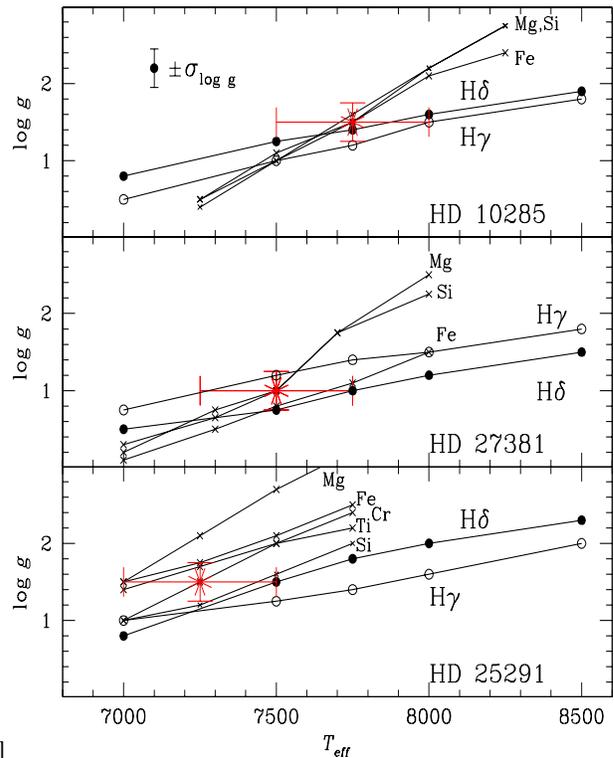} 
            \caption{Feature trends with variations in gravity and temperature.
The asterisk indicates the adopted pair $T_{\rm eff}$-log $g$ for subsequent
element abundance analysis. The Balmer line loci were obtained by fitting the observed
lines with the Kurucz's (1993) models. The uncertainty in log $g$ for a given temperature attained in this process is 
indicated by the error bar in the top panel.} 
            \end{figure}

      \section {Derived elemental abundances}\label{specana}   

      In this section, we present our abundance results for the individual 
    stars. In the abundance tables 4 and 6 the solar abundances
   were taken from  the work of Grevesse, Noels \& Sauval (1996) except
  for carbon  and oxygen that were 
  taken from Allende Prieto, Lambert \& Asplund (2002; 2001) respectively. The 
   number of lines included in the calculations is represented by n.

      \subsection{HD 10285} 

  HD 10285 is considered a high galactic
   latitude low-mass supergiant candidate by Bidelman 
  (1990) based on its high Str\"omgren $c_1$ index (1.62).  
  Our interest in this object has been  aroused by its significantly
  large radial velocity 
 ($-58$ km~s$^{-1}$). As can be seen from Table 4 our analysis covers a
  large number of elements  based on a large number of lines for many elements.
   Although the derived iron deficiency [Fe/H] $\sim -0.3$ is not very large, 
   it is  a high precision estimate. 
   The star also shows noticeable relative enhancement of Na and S relative 
to Fe.  Non-LTE corrections for Na abundances at the temperature of 7750~K
   for the lines employed by us is $\sim -0.16$  as tabulated by 
    Takeda \& Takada-Hidai (1994). 
   Hence the Na enhancement reported in Table 4 is not caused 
    by the neglect of non-LTE effects but is a real effect. 
    The enhancement of Na relative to Fe is possibly caused by the 
    mixing of the NeNa-cycle products from the hydrogen burning region.   
    We do not have evidence of evolution beyond the red giant stage
    for this star.

           \begin{table*} 

           \caption{Elemental abundances of hotter stars } 
      \scriptsize{ 

      \begin{center} 
      \begin{tabular}{lcccccc}   
      \noalign{\smallskip}                                                               
      \hline       
      \noalign{\smallskip} 

      \noalign{\smallskip} 
      \multicolumn{1}{c}{Spe}& 
      \multicolumn{1}{c}{}& 
      \multicolumn{1}{c}{HD 10285}& 
      \multicolumn{1}{c}{HD 25291}& 
      \multicolumn{1}{c}{HD 27381}& 
      \multicolumn{1}{c}{HD 137569}& 
      \multicolumn{1}{c}{HD 172324}\\ 
      \noalign{\smallskip} 
           \cline{3-3} \cline{4-4} \cline{5-5} \cline{6-6} \cline{7-7} 
      \multicolumn{1}{c}{cies}& 
      \multicolumn{1}{c}{$log \epsilon_{\odot}$}& 
      \multicolumn{1}{c}{[X/H], n, [X/Fe]}& 
      \multicolumn{1}{c}{[X/H], n, [X/Fe]}& 
      \multicolumn{1}{c}{[X/H], n, [X/Fe]}& 
      \multicolumn{1}{c}{[X/H], n, [X/Fe]}& 
      \multicolumn{1}{c}{[X/H], n, [X/Fe]}\\ 
      \noalign{\smallskip}   
      \hline   
      \noalign{\smallskip}     

           C I& 8.39 & $-0.27\pm 0.13$, 3, $+0.04$ & $-0.71\pm0.17$, 5, $-0.38$& $-1.04$, syn,$-0.37$  & . . . & . . . \\ 

      C II& 8.39 & . . . & . . . & . . . & $-0.2$,syn,$+2.8:$ & $-0.04$,syn,$+0.5$  \\ 
    &  & . . . & . . . & . . . & $\bf {-0.6}*$,syn,$+2.4:$ & $\bf {-0.3}*$,syn,$+0.2$  \\ 

           N I& 7.83 & . . . &. . . &$+0.50$, syn,$+1.20$ & $+0.8\pm0.06,syn, +3.8:$ & $+0.7\pm0.16, syn,+1.2$   \\ 
      &  & . . . &. . . &$\bf {+0.2}*$,syn,$+0.9$ & $\bf {+0.2}*$,syn,$+3.2:$  & $\bf {+0.1}*$,syn, $+0.6$    \\ 

           O I& 8.69 & $-0.00\pm 0.03, 2,+0.30$ & $-0.18\pm0.01, 2,+0.15$ & $-0.36\pm0.03,  3,-0.3$ & $+0.3\pm0.16, 9, +3.3:$ 
           &  $+0.2\pm0.07, 5,+0.8$  \\ 

      & &  &  & & $\bf {+0.1}*$,syn,$+3.1:$   & $\bf {-0.1}*$,syn,$+0.4 $ \\ 

           Ne I& 8.09 & . . . &. . . & . . .   & $+0.2\pm0.14, 3, +3.2:$ &   $+0.4\pm0.16, 5,+0.9$   \\ 

            Na I& 6.33 & $+0.11\pm 0.04, 2,+0.41$ & $+0.17\pm0.05, 4,+0.50$ & $-0.13\pm0.04, 3,+0.54$ & . . .  & . . . \\ 

           Mg I& 7.58 & $-0.19\pm 0.20, 5,+0.11$ & $-0.26\pm0.12, 4,+0.07$ & $-0.76\pm0.01, 2,-0.09$ & . . .   &  $-0.4\pm0.21, 3,+0.1$  \\ 

           Mg II& 7.58 & $-0.21\pm 0.18, 3, +0.10$ &$-0.48\pm 0.10, 2,-0.15$  & $-0.74,\pm0.08, 2,-0.07$ & $-2.8, 1,+0.2 $ 
           &  $-0.4, 1, +0.1$   \\ 

           Si I& 7.55 & $-0.37\pm 0.17, 3,-0.06$ & $+0.08\pm0.09, 9,+0.42$ & $-0.20\pm0.03, 2,+0.47$ & . . .  & . . . \\ 

           Si II& 7.55 & $-0.39\pm 0.05, 2,-0.09$ & $+0.23\pm0.13, 2,+0.56$ & $-0.10\pm0.25, 2,+0.57$ & $-2.4\pm0.17, 2,+0.6 $ 
           &  $-0.1\pm0.10, 7,+0.4 $   \\ 
            
           Si III& 7.55 & . . .& . . . & . . . &. . . & $+0.0,1,+0.5 $\\ 

           S I& 7.21 & $+0.04\pm 0.02, 2,+0.35$ & $+0.08\pm 0.08, 3,+0.41$ & $-0.30,  0.21, 2,+0.37$ & . . . &. . . \\ 

           S II& 7.21 & . . . & . . . & . . . & $-0.1\pm0.20,15, +0.6 $  & $+0.1\pm0.18, 9,+0.6$ \\ 

           Ca I& 6.36 & $-0.18\pm0.15, 9,+0.13$ & $-0.18\pm 0.12,22,+0.15$ & $-0.59\pm0.09, 7,+0.09$ & . . . & . . . \\ 

           Ca II& 6.36 & $-0.19 $, 1, $+0.12$ & $-0.26$      , 1,$+0.07$ & . . . & . . .   
           & $-1.5\pm0.05,2,-1.0$ \\ 

           Sc II& 3.10 & $-0.07\pm 0.10,10,+0.24$ & $+0.01\pm0.18,11,+0.34$ & $-0.39\pm0.14, 7, +0.28$& . . . 
           & $-0.6, 1,-0.1$\\ 

           Ti I& 4.99 & . . . & $-0.25\pm0.13, 4,+0.08$ & . . . &. . . & . . . \\ 

           Ti II& 4.99 & $-0.30\pm 0.15,27,+0.01$ & $-0.34\pm0.16,17,-0.01$ & $-0.57\pm0.19,25,+0.10$ & . . . 
           & $-0.4\pm0.15,22,+0.1$ \\ 

           V  II& 4.00 & $-0.03,1,+0.28$& $-0.28\pm0.28,2,-0.05$& . . . & . . . & . . .   \\ 

           Cr I& 5.67 & $-0.27\pm 0.25 , 4,+0.04$ & $-0.23\pm0.12,10,+0.10$& $-0.59, 1,+0.09$ & . . . 
           & . . .  \\ 

           Cr II& 5.67 & $-0.24\pm 0.14,19,+0.07$ & $-0.25\pm0.11,14,+0.09$ & $-0.44\pm0.17,23,+0.23$ & . . . 
           & $-0.5\pm0.15,13,-0.0$ \\ 

           Mn  I& 5.39 & $-0.27, 1,+0.04$ & $-0.31\pm0.16, 5,-0.02$ & $-0.49\pm0.15, 3,+0.18$ & . . . 
            & . . . \\ 

           Mn  II& 5.39 & $-0.12\pm 0.27, 2,+0.11$ & . . .          & . . . & . . . 
           & . . . \\ 

           Fe  I& 7.52 & $-0.30\pm 0.17$,87,....... & $-0.28\pm 0.12,136,......$ & $-0.71\pm0.12,33, .......$ & . . . 
           & . . .  \\ 

           Fe  II& 7.52 & $-0.32\pm 0.13$,28,....... & $-0.38\pm 0.12,29,......$ & $-0.63\pm0.11,24, .......$ 
           &  $<-3.0, syn, $ & $-0.5\pm0.14,36, .......$\\ 

           Co  I& 4.91 & . . . & $-0.42, 1, -0.09$ & . . .  & . . . & . . . \\ 

           Ni  I& 6.25 & $-0.17\pm 0.09, 6,+0.14$ & $-0.24\pm0.10,17,+0.09$ & . . . & . . . & . . .\\ 
           Ni  II& 6.25 & $-0.37\pm 0.22, 2,-0.06$ & . . .  & . . . & . . .  & . . . \\ 

           Zn  I& 4.63 & $-0.36\pm0.23, 2,-0.06$ & . . .  & . . . & . . . & . . . \\ 
           Sr  II& 2.90 & $-0.06, 1,+0.25$ & . . .  & $-1.03, 1,-0.36$  & . . . & . . . \\   
           Y  II& 2.24 & $-0.30\pm 0.07, 4,+0.01$ & $-0.17\pm0.21, 4,+0.16$ & $-0.51\pm0.18, 5,+0.16$& . . . &. . . \\ 
           Zr II& 2.60 & $-0.34, 1,-0.03$ &$-0.33\pm0.01, 2,+0.00$ & $-0.80, 1,-0.13$ & . . . & . . . \\ 
           Ba II& 2.13 & $-0.11\pm 0.13, 3,+0.20$ & $+0.01\pm0.05, 2, +0.34$& $-0.61\pm0.12, 4,-0.06$ & . . .  & . . .  \\ 
           Ce II& 1.55 & . . . & $-0.12\pm0.11, 2,-0.21$& . . . &. . .  & . . .  \\ 
           Eu II& 0.51 & . . . & $-0.47, 1,-0.14$& . . . &. . . & . . . \\ 
                 \noalign{\smallskip}   
                 \hline   
                 \noalign{\smallskip} 
      \end{tabular} 
      * Non-LTE corrected estimates. 
      \end{center} 
      } 
               \end{table*} 

      \normalsize 

      \subsection{HD 25291} 

       This star has been
     included in a number of investigations and the estimated temperatures
    lie in the 6700~K (Andrievsky et al. 2002) to 7600~K (Venn 1995a,b) range. 
   Venn carried out a comprehensive abundance analysis covering many elements.
   Our analysis employs an even larger number of lines and extends to heavier elements.
   The carbon abundance derived by us is systematically smaller 
   than the value derived by Venn (1995b) but the lines used by us are
   different. For silicon and oxygen we find good  
    agreement with the values obtained by Venn. 
    We find a relative enrichment of Na, Si and S, similar to that 
      found in HD 10285. The abundance of s-process elements is essentially
   solar. We do not consider this object as a post-AGB star. 

      \subsection{HD 27381} 

    Although the SED generated using observed optical and IR 
         color bands shows larger than stellar component fluxes at 
         12, 25 and 60 $\mu$, the error bars are too large to report 
          IR excess. At 100$\mu$ the excess IR flux is larger 
    than the measurement errors.  We have derived carbon 
   abundance by synthesizing the the 4765-4780\AA~ spectral region containing 
   CI lines as shown in  Figure 2. Nitrogen abundances were derived
    using N~I lines in the 7420-7470\AA~ region.
     As can be seen from the abundances presented in Table 4,
      this is a metal-poor star. Although [Fe/H] of $-0.7$ is not very
      large, it is quite significant.  Since the derived [Fe/H] 
       is  very similar to the mean metallicity of thick disk population 
   (Gilmore et al. 1995), we chose to compare elemental abundances for 
 this star with those derived for thick disk stars by Prochaska et al. (2000).
           These authors give the mean [X/Fe] for a large number 
           of elements in their Table 19. Although 
       HD 27381 shows enhancement of some $\alpha$-elements like O, Si and S 
      seen in the thick disk  stars, it does not show relative enrichment of 
            Mg and Ti. The observed [Si/Fe] and [S/Fe] are larger than 
           their corresponding values in Prochaska et al.'s (2000) sample, 
  and  relative enhancement of Al is not seen, 
           although [Ca/Fe] agrees within the errors of the estimates.
            On the other hand, we find an indication of 
           evolutionary effects in the form of enhanced N and Na abundances
           and carbon deficiency. 
           A non-LTE correction of $-0.4$ is estimated by
          Venn(1995b) for the C~I lines at the temperature of 7500~K,
           although this estimate was made for lines in 7100\AA~ region.
          Hence the  non-LTE [C/Fe] would be even smaller than the [C/Fe]
        of $-0.4$  given in Table 4. For N~I lines used in the present work at 7500~K 
           the non-LTE corrections are estimated as  $-0.3$ indicating
          a non-LTE corrected [N/H] of $+0.2$ and [N/Fe] of $+0.9$. Hence we have
          a very distinct signature of CN processed material on the surface
          of this object.
  This star resembles HD 107369 as described by 
            Van Winckel (1997) in some respects. Its temperature and 
   hydrogen lines profiles are similar, both objects are metal-poor with 
           [Fe/H] of $-1.0$ and $-0.7$ for HD 107369 and HD 27381 respectively,
     and  show a relative    carbon deficiency and nitrogen
           enhancement. These two objects show in a milder form, the
  abundance pattern  reported for hot post-AGB stars by Conlon et al. (1993).

            \begin{figure}
            \includegraphics[width=8.cm,height=8.cm]{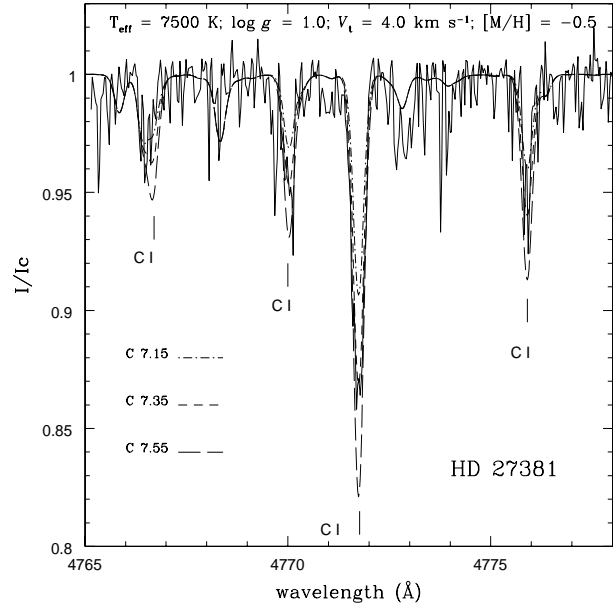} 
            \caption{Agreement between the synthesized and observed spectrum 
            for carbon lines. The best fit is obtained for 7.35.} 
            \end{figure} 

      \subsection{HD 137569} 

The reddening-free indices [$m_1$] = 0.062, [$u-b$] =0.939 and [$c_1$] =0.707 when used with
synthetic colors for model atmospheres of Kurucz(1993) led to a preliminary estimates
of temperature and gravity. We have used Str\"omgren photometry published in
Mermillod(1998) and Newell \& Graham (1976). The indices [$u-b$] and [$c_1$] are the measure
of Balmer discontinuity, which for this temperature range is known to be a good 
temperature indicator, while it is not sensitive to gravity.
 This led us to a starting $T_{\rm eff}$
 of 12000~K. A similar value is reported by Danziger \& Jura(1970)
 These authors derive log $g$ of 2.3 
 from the intersection of the H$_{\gamma}$ profile and Balmer jump loci.
More recently Behr (2003) estimated rotation velocity and atmospheric parameters
by comparing the observed spectrum with the synthesized one.
 Behr estimated $T_{\rm eff}$
of 12073$\pm$400~K, log $g$ of 2.38$\pm$0.4, $\xi$ of
 0.0$\pm$0.8 kms$^{-1}$ and $v$~sin~$i$ 18.5 km~s$^{-1}$.

      \begin{figure} 
       \includegraphics[width=8cm,height=8cm]{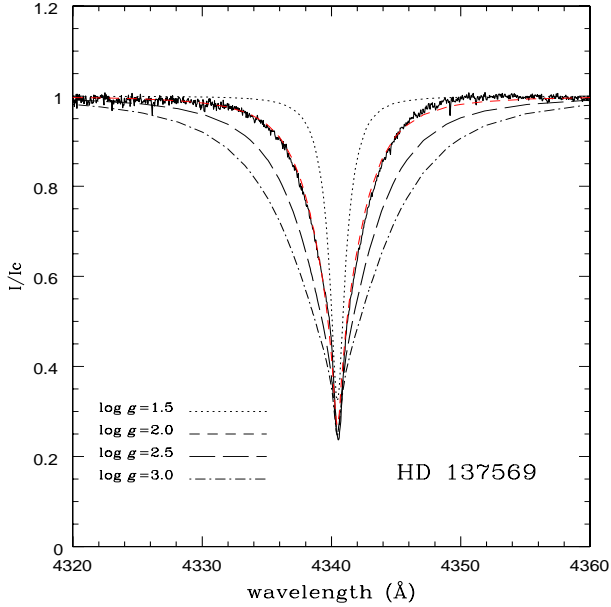}   
    \caption{H$\gamma$ profile of HD 137569. The models correspond to
 T$_{\rm eff}$ 12000~K and gravities 1.5, 2.0, 2.5 and 3.0. The best fit is
  obtained for log $g$ = 2.0.} 
       \end{figure} 
 
We had obtained one spectrum of the star in 2000
 at the OHP and were
intrigued by its appearance. Although the H~I and He~I lines were
 similar to that of a late B supergiant, the C~II, Si~II, Mg~II and
 Fe~II lines were extremely weak. Another spectrum using the 2.7m 
telescope of McDonald Observatory was obtained with a 
resolution of 60000 which enabled a spectral coverage of 3700\AA~ to 1$\mu$.
A comparison of observed H$_{\gamma}$ and H$_{\delta}$ profiles with those
synthesized using $T_{\rm eff}$=12000~K and a range
 of gravities lead to log $g$=2.0. We have used the models by
 Jeffery, Woolf \& Pollaco (2001) which extend to much lower gravities
than those of Kurucz(1993). These models could also give a good match to the He~I
line at 4388\AA~ which is also gravity sensitive and to
 He~I lines at 5015, 5047\AA~ which are  very sensitive to temperature.
 We have shown in Figure 3 the comparison between computed profiles for 
 different gravities and the observed H$_{\gamma}$ profile.

 The estimated temperatures and
gravities therefore have accuracies of $\pm$500~K and $\pm$0.5 dex. The N~I lines
have a wide range in equivalent width and therefore were used to derive $\xi$ of 
7 km~s$^{-1}$. Our derived abundances for different elements are presented in Table 4.
 The agreement between the synthesized and observed spectrum for the spectral region
  covering the C, N, O region is presented in Figure 4. 
 The figure also contains the spectrum synthesized using the adopted atmospheric
 parameters and solar composition (shown as large dots in Fig. 4). It is obvious
 that the star is highly deficient in Fe. The lines of Fe II at 6147.7, 6149.3, 7462.4
 are nearly absent in the observed spectrum.

        \begin{figure*} 
       \includegraphics[width=18cm,height=11.5cm]{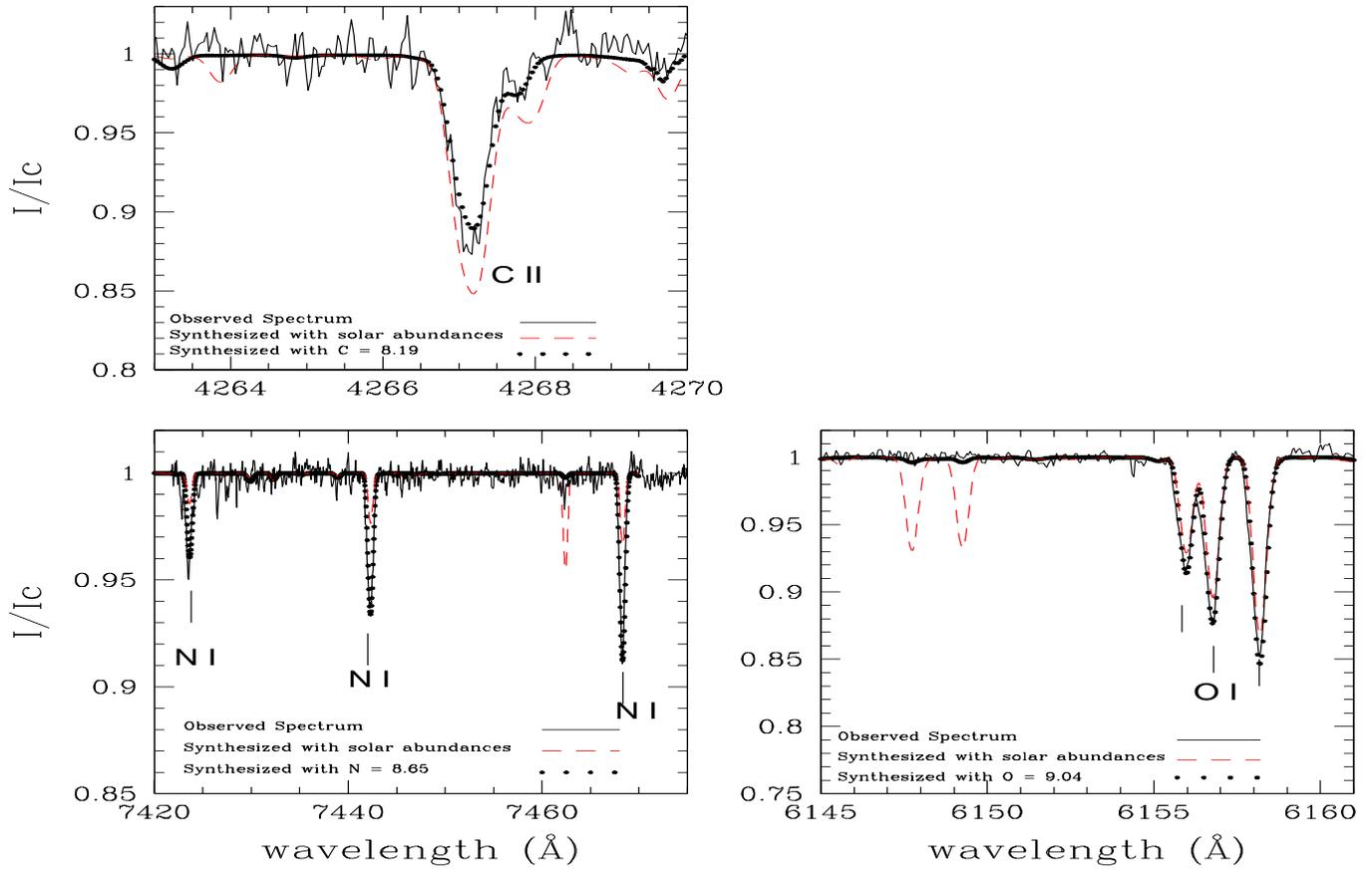}   
      \caption{The agreement between  the synthesized and observed spectrum for  spectral 
      regions containing  C, N and O lines for HD 137569} 
       \end{figure*}

At the temperature of 12000~K, non-LTE effects are expected to be large. Przybilla
 et al. (2001, 2000) have carried out a very extensive non-LTE analysis of C,N,O elements for stars in the 9000 to 12500~K temperature range. The star $\beta$ Ori studied
by them has atmospheric parameters ($T_{\rm eff}$=12500~K, log $g$=1.8) very similar to
those of HD 137569 and the investigation employs the same lines. Therefore the non-LTE
correction calculated for $\beta$ Ori may be used for HD 137569. After applying
a non-LTE correction of $-0.4$ for C~II, $-0.6$ for N~I and $-0.30$ for O~I lines
 we present the non-LTE corrected estimates in bold in Table 4.

The derived deficiency of C accompanied by N enhancement indicates that the
 CN processed material is brought to the surface via dredge-up.
 The carbon deficiency found for HD 137569 is not as large as 
in other hot post-AGB stars. A more remarkable feature is 
 the total absence of Fe~II lines. We could only get an upper estimate of $-3.0$ for
[Fe/H]. The weak Si~II and Mg~II lines also led to
 [Si/H] and [Mg/H] of $-2.4$ and $-2.8$
respectively. On the other hand the S~II and Ne~I lines indicate near solar abundance.
These abundances are derived using a sufficiently large number of lines
 and hence are quite robust.
The selective depletion of easily condensable elements with higher condensation
temperatures has been observed in a large number of post-AGB stars. Well known
examples are HR 4049 (Lambert et al. 1988) and HD 44179 (Van Winckel 1995). A similar
phenomenon has also been found in many RV Tau stars (Giridhar et al. 2005, Maas 2002
  and references therein). The elements S and Zn have a lower condensation
 temperature and are generally
unaffected by this effect. In fact these elements are considered better
 metallicity indicators. However, S and Zn abundances are not available 
 for many hot post-AGB stars.

      \begin{figure} 
       \includegraphics[width=8cm,height=8cm]{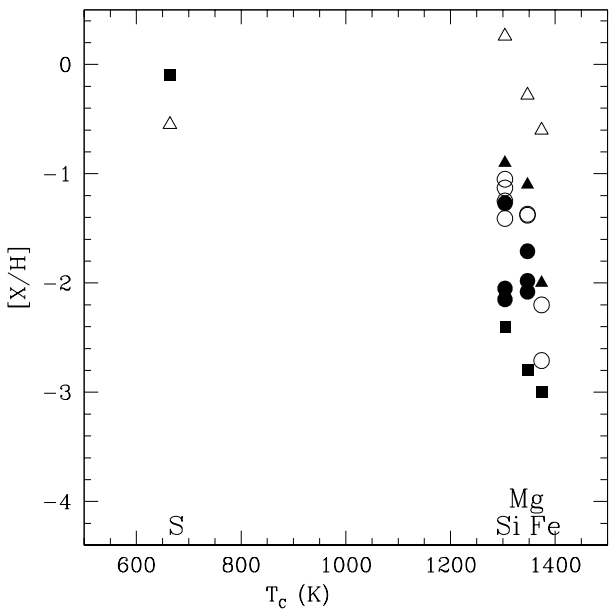}   
      \caption{ Elemental abundance plotted as a function of condensation 
temperature. Filled symbols represent relatively cool stars
 and open symbols represent hotter stars. 
      Triangles  represent PNe stars following the same convention.
 The post-AGB stars are shown as circles. 
       Filled squares are the values for HD 137569 for which the depletion
 due to condensation is quite evident.} 
       \end{figure} 

Although this phenomenon is well studied in post-AGB stars
 of intermediate temperatures,
its presence in hot post-AGB stars is not yet established.
  We are only aware 
of LS IV $-4^{o}.01$ and LB 3193 showing incomplete suggestions of
 selective depletion of refractory elements. We have compared the abundances of
 HD 137569 with other hot post-AGB
candidates and two PNe given in Table 5.
 We have plotted in Figure 5 the
observed abundances of the stars included in Table 5 as a function of
the most recent condensation  temperatures presented by Lodders (2003).
 It might have been more instructive to plot the abundances of these 
elements relative to S or Zn but these abundances are not available
 for many post-AGB stars. The depletion of refractory elements is 
 an attractive but yet to be substantiated possibility 
 for stars cooler than 13000~K. Efforts are required to measure sulphur
abundances in hot post-AGB candidates.
 The large depletion only of refractive elements gives strong
  support to the post-AGB candidature of HD 137569.
The explanation of these abundance peculiarities 
found in post-AGB stars requires a site of dust-gas separation. The suggestion
 of dust-gas separation occurring in a circumbinary disk is 
gaining more support with the increase in detected binaries
among post-AGB stars. HD 137569 is known to be a single lined spectroscopic
binary with a period of 529.8 days and a systemic velocity
 of $-45$ km~s$^{-1}$ as 
reported by Bolton \& Thomson(1980). This  might strengthen our suggestion
that HD 137569  is a post-AGB star. Lack of IRAS fluxes might be caused by 
a longer transition time. 
 The derived temperature and gravity places the star very near the
post evolutionary track of Sch\"onberner (1983) for a mass of 0.546 M$_{\odot}$.
The C,N,O abundances  
of HD 137569 resemble with those for
 PN LS IV $-12^{o}.111$ (Conlon et al. 1993).

The UV spectra of this object would be extremely useful for the better coverage 
of Fe-group elements.

      \begin{table*}[!t] 
           \caption{ Abundances for HD 137569 compared with 
           other post-AGB and PNe stars}                                                                 
           \begin{center} 
           \begin{tabular}{lcccccccc} 
           \noalign{\smallskip}                                                               
           \hline       
           \noalign{\smallskip}                                                               
           \noalign{\smallskip} 
           \multicolumn{1}{l}{Star}&   
           \multicolumn{1}{c}{$T_{\rm eff}$}&   
           \multicolumn{1}{c}{[C/H]}&   
           \multicolumn{1}{c}{[N/H]}& 
           \multicolumn{1}{c}{[O/H]}& 
           \multicolumn{1}{c}{[Mg/H]}& 
           \multicolumn{1}{c}{[Si/H]}& 
           \multicolumn{1}{c}{[S/H]}& 
           \multicolumn{1}{c}{[Fe/H]}\\ 
                      \noalign{\smallskip}   
                      \hline   
                      \noalign{\smallskip}   
           LS IV $-4^{o}.01$$^1$& 11000&$<-1.21$& & &$-1.98$&$-2.05$& & \\ 
           LS IV $-4^{o}.01$$^8$& 10000&       & & &$-2.78$&$-1.68$& &$-1.71$ \\ 
           {\bf HD 137569}$^2$& 12000&$-0.60$&$+0.20$&$+0.05$&$-2.80$&$-2.40$&$-0.10$&$ <-3.0$\\ 
           LB 3193$^1$& 12900&$<-1.91$& & &$-2.08$&$-2.15$& & \\ 
           PG 1323-086$^3$& 16000&$<-2.25$&$-0.98$&$-0.77$&$-1.71$&$-1.37$& & \\ 
           PG 1704+222$^3$& 18000&$-1.40$&$-0.89$&$-0.85$&$-1.37$&$-1.13$& & \\ 
           PHL 174$^{4}$& 18200&$<-2.31$&$-1.11$&$-0.85$&$-1.38$&$-1.05$& & \\ 
           Barnard 29$^{5}$& 20000&$-2.16^4$&$-0.61$&$-1.15^1$& &$-1.25^5$& &$-2.2^5$\\ 
           BD +33$^0$ 2642$^6$& 20200&$-0.93$&$-0.56$&$-0.62$&$-1.1$ &$-0.9$& & \\ 
           LB 3219$^1$& 21400&$-1.71$&$-0.31$&$-1.15$& $-0.58$&$-0.95$& & \\ 
           ROA 5701$^2$& 24000&$-2.57$&$-1.04$&$-0.78$& &$-1.41$& &$-2.71$\\ 
           LS IV $-12^o.111$$^{1}$& 24000&$-1.71$&$-0.11$&$+0.05$&$-0.28$&$+0.05$&$-0.6$&$-0.41$\\ 
           LS IV $-12^o.111$$^{7}$& 24000&$-0.63$&$+0.29$&$+0.05$&$-0.28$&$+0.08$&$-0.5$& \\ 
                      \noalign{\smallskip}   
                      \hline   
                      \noalign{\smallskip}   
           \end{tabular} 
           \end{center} 

           References: (1) McCausland et al. (1992), 
           (2) this work, (3) Moehler \& Heber (1998), 
           (4) Conlon et al. (1991), (5) Moehler et al. (1998), 
           (6) Napiwotzki et al. (1994), (7) Conlon et al. (1993), 
           (8) Mooney et al. (2002) 
           \end{table*}

      \subsection{HD 172324} 

   HD 172324 is a hot star at high galactic latitude, showing emission
     components in the hydrogen line profiles. It is a  high radial velocity
     star and velocities around  $-110$ km~s$^{-1}$ are seen.
     We observed the stars in 1995, 1999, 2000 and 2003 and found
  radial velocity variations larger than the measurement errors 
   The measured heliocentric radial 
  velocities are  given   in Table 2. The star is listed as a
 possible variable star in the catalog of Rufener (1981). 

     We have obtained high resolution spectra of this object at different
     epochs and found very interesting variations in H$_{\alpha}$
 and H$_{\beta}$ as displayed in Figure 6.
The H$_{\alpha}$ profile is seen  mostly in emission.
H$_{\beta}$ has a sharp  central emission, rising above the continuum 
level on March 2003, and a broad absorption line.
   The spectrum contains a large number of He I lines.
 We have used the excitation and ionization equilibrium of Fe,
 Mg and Si to derive $T_{\rm eff}$=10500~K, log $g$=2.5 and
 $\xi$ = 7.0 km~s$^{-1}$.
   The presence of a strong O~I triplet at 7774\AA~  also supports the derived
   low gravity for this star. 
  The lines used for C,N,O abundances are the same as those  used for HD 137569.
    We have derived almost solar values [C/H]=$-0.04$,
   [N/H]=$+0.71$ and [O/H]=$+0.25$ for this star.
   At the temperature of 10500~K,
 departure from non-LTE conditions  are likely to be significant.
  Using the non-LTE correction for C~II feature tabulated by
  Przybilla, Butler \& Kudritzki (2001)
 for stars in the temperature range 9000
  to 12500~K, we have estimated a non-LTE correction of $-0.3$ hence the
 corrected [C/H]=$-0.34$.
 Similarly, using non-LTE corrections of $-0.6$  for N~I (Przybilla \& Butler  2001)
  that of  $-0.3$ for O~I (Przybilla et al. 2000) we have 
  derived the corrected values  of [N/H]=$+0.11$ and [O/H]=$-0.05$, i.e. nearly 
	solar values for HD 172324. These abundances are very similar to 
   the C,N,O estimates derived for a sample of  population I B type
   supergiants  (Gies \& Lambert 1992).
   However, consistently large radial velocity derived at 4 epochs
   and [Fe/H]=$-0.5$ makes it a very intriguing object.

  We were surprised by the absence of the Na I 8195\AA~ feature in our
   spectrum.  For $\eta$ Leo ($T_{\rm eff}$ $\sim$ 10200~K), Takeda \&
   Takada-Hidai (1994) have reported a strength of 30~m\AA. Since 
   the temperature of HD 172324 is very similar,
 the non-detection of the above Na feature would imply that [Na/H]
   $\le -0.8$. For a sample of A-F supergiants, these authors have
   reported an overabundance of Na possibly caused by the mixing of
   NeNa-cycle products from the hydrogen burning zone.
   Absence of Na overabundance  indicates that the NeNa-cycle
   products have not been brought to the surface. On the other hand,
   Ne~I lines are very strong, indicating [Ne/H]=$+0.4$.

   It appears that similar to A supergiants studied by
  Venn(1995b) the evolution of HD 172324 avoided extensive
   mixing at the red giant phase  by evolving directly from the main sequence
   to their present position. Relative to iron, Ne, Si, and S show  
  the mild enrichment normally seen in metal-poor objects. 

  \begin{figure}
  \includegraphics[width=8.cm,height=8.cm]{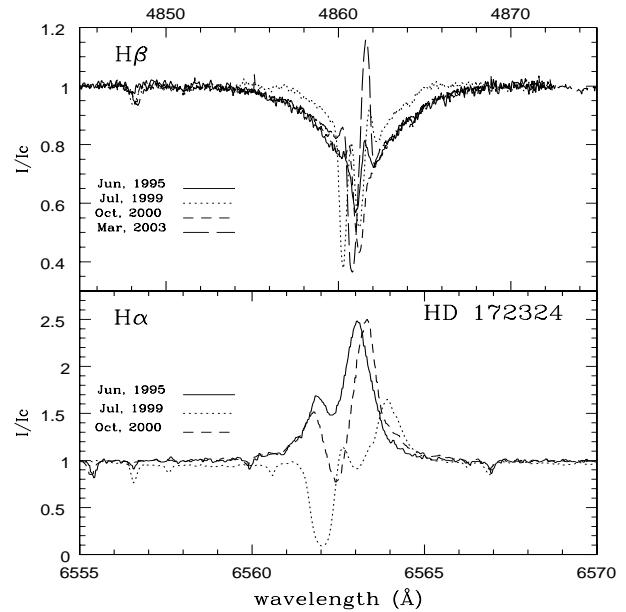}
  \caption{H$\alpha$ and H$_\beta$ variations in HD172324.}
  \end{figure}

   The radial velocity of this star shows significant variation
   as can be seen   in Table 2. The same is true for hydrogen
 line profiles where the strength of emission components show large
  variations. Our high resolution spectrum of this
   star shows distinct doubling for O~I lines in the 7774\AA~
  triplet components.  We therefore believe it may be a pulsating variable
  or a binary star.

      \scriptsize 

           \begin{table*} [!t]

           \caption{ Abundances of cooler stars } 

      \scriptsize{ 

      \begin{center} 
      \begin{tabular}{lcccc}   
      \noalign{\smallskip}                                                               
      \hline       
      \noalign{\smallskip} 

      \noalign{\smallskip} 
      \multicolumn{1}{c}{Species}& 
      \multicolumn{1}{c}{}& 
      \multicolumn{1}{c}{HD 10132}& 
      \multicolumn{1}{c}{HD 12533}& 
      \multicolumn{1}{c}{HD 159251}\\ 
      \noalign{\smallskip} 
           \cline{3-3} \cline{4-4} \cline{5-5} 
      \noalign{\smallskip} 
      \multicolumn{1}{c}{}& 
      \multicolumn{1}{c}{$log \epsilon_{\odot}$}& 
      \multicolumn{1}{c}{[X/H], n, [X/Fe]}& 
      \multicolumn{1}{c}{[X/H], n, [X/Fe]}& 
      \multicolumn{1}{c}{[X/H], n, [X/Fe]}\\ 
      \noalign{\smallskip}   
      \hline   
      \noalign{\smallskip}     

           C I& 8.39 & $-0.43\pm 0.11, 2,-0.39$ & . . . & . . .  \\   
           O I& 8.69 & $-0.20, 1,-0.16$ & . . . & . . .  \\ 
           Na I& 6.33 & $-0.04\pm 0.19, 4,+0.01$ & $+0.14\pm0.16, 2,+0.20$ & $-0.10\pm0.03, 2,-0.06$\\ 
           Mg I& 7.58 & $-0.18\pm 0.37, 3,-0.14$ & $-0.24, 1,-0.18$ &$-0.67, 1,-0.63$ \\ 
          Al I& 6.47 & $-0.35\pm0.14, 3,-0.31$& $-0.15, 1,-0.09$& $-0.21\pm0.09, 2,-0.17$  \\ 
            Si I& 7.55 & $+0.03\pm0.16,16,+0.08$ & $+0.14\pm0.25, 8,+0.20$ & $+0.04\pm0.11,20,+0.09$\\ 
           S I& 7.21 & $+0.15\pm 0.11, 2,+0.20$ & . . . & . . .\\ 
           K I& 5.12 & . . .  & $+0.21, 1,+0.27$ & . . . \\ 
           Ca I& 6.36 & $+0.05\pm0.20,11,+0.10$ & $-0.20\pm0.19,10,-0.14$&$-0.21\pm0.25,25,-0.17$ \\ 
           Sc I& 3.10 & . . .  &$-0.12\pm0.16,7,-0.06$&$-0.16\pm0.15,10,-0.11$\\ 
            Sc II& 3.10 &$+0.36\pm 0.17, 9,+0.41$ & $-0.03\pm0.20, 5,+0.03$ &$+0.10\pm0.32, 9,+0.15$\\ 
            Ti I& 4.99 & $-0.07\pm0.16,20,-0.03$ &$-0.19\pm0.17,38,-0.13$ &$-0.13\pm0.21,81,-0.09$\\ 
             Ti II& 4.99 & $+0.04\pm0.10, 5,+0.09$ & $-0.22\pm0.25, 4,-0.16$ &$-0.07\pm0.09, 5,-0.03$ \\ 
            V  I& 4.00 & $-0.05\pm0.24,15,-0.01$ & $+0.01\pm0.24,31,+0.07$ &$0.03\pm0.20,35,+0.08$\\ 
            V  II& 4.00& . . . & . . . & $+0.17, 1,+0.22$ \\ 
             Cr I& 5.67 & $-0.03\pm 0.19,19,+0.02$&$-0.16\pm0.20,22,-0.10$& $-0.07\pm0.22,35,-0.03$ \\ 
             Cr II& 5.67 & $-0.09\pm0.20,10,-0.05$ & $+0.04\pm0.25, 2,+0.10$ & $+0.07\pm0.16, 7,+0.12$ \\ 
             Mn  I& 5.39 &$+0.20\pm0.20, 5,+0.25$ & $+0.01\pm0.21, 6,+0.07$ & $+0.02\pm0.31,16,+0.07$\\       
          Fe  I& 7.52 & $-0.01\pm 0.14$,94,....... & $-0.08\pm0.18,132,.......$&$-0.03\pm0.16,253,.......$\\ 
             Fe  II& 7.52 & $-0.08\pm 0.12$,19,....... &$-0.04\pm 0.22,11,.......$&$-0.06\pm0.12,18,.......$\\ 
             Co I& 4.92&$-0.07\pm 0.18, 6,-0.03$ &$-0.02\pm0.25,11,+0.04$&$+0.17\pm0.24,23,+0.22$\\ 
            Ni  I& 6.25 & $-0.07\pm0.17,27,-0.03$ & $-0.05\pm0.17,38,+0.01$&$-0.04\pm0.24,88,+0.01$\\ 
              Cu  I&4.26 & . . . & . . .  &$+0.41\pm0.33,3,+0.46$ \\ 
              Zn  I& 4.60 & $-0.14\pm0.04, 2,-0.10$&$-0.05, 1,+0.01$ & $-0.14\pm0.20,2,-0.09$ \\ 
              Sr  I& 2.90 & . . .  & . . .  & $-0.21,1,-0.16$ \\ 
              Y I& 2.24 & . . . & $-0.24\pm0.16, 2,-0.18$ & . . .\\ 
             Y  II& 2.24 & $-0.05\pm0.13, 4,+0.00$ & $-0.12\pm0.15, 3,-0.06$ & $-0.06\pm0.21, 3,-0.01$\\ 
             Zr I& 2.60 &. . . &. . . & $-0.10\pm0.12, 6,-0.05$ \\ 
             Zr II& 2.60 & $+0.12\pm0.04,2,+0.17$ &$-0.26\pm0.16, 9,-0.20$ & $+0.29, 1,+0.34$ \\ 
             Mo II& 1.92 & . . .&$+0.04\pm0.10, 2,+0.10$ & $-0.01, 1,+0.04$ \\ 
             Ba II& 2.13 & $+0.19\pm0.09,2,+0.24$ & $+0.23, 1, +0.29$& $+0.02\pm0.13, 2,+0.07$\\ 
             La II& 1.22&$-0.17\pm0.04, 3,-0.13$ & $+0.13\pm0.18, 2,+0.19$&$-0.18\pm0.14, 3,-0.13$\\ 
             Ce II& 1.55&$+0.02\pm0.20, 3,+0.07$ & $-0.01\pm0.15, 4,+0.05$&$-0.09\pm0.21, 4,-0.04$ \\       
             Pr II& 0.71 & . . . &$+0.34, 1,+0.40$ &$+0.16\pm0.10, 2,+0.21$ \\ 
             Nd II&1.50 &$-0.11\pm0.17, 6,-0.07$ &$0.12\pm0.54, 5,-0.06$  &$-0.16\pm0.18, 5,-0.11$ \\       
             Sm II& 1.00 & . . . & . . . &$+0.01\pm0.30, 2,+0.06$  \\ 
            Eu II& 0.51 &$-0.04\pm0.07, 2,+0.01$ & . . .&$+0.17,1,+0.22$  \\ 
                 \noalign{\smallskip}   
                 \hline   
                 \noalign{\smallskip} 
      \end{tabular} 
      \end{center} 
      } 
               \end{table*} 

      \normalsize 

      \subsection{Cool stars} 

      These stars were included in our sample due to IR flux detection and 
      their high galactic latitudes. We could measure a very large number of lines 
      for each element and for many elements lines of two stages of
      ionization could be used.  We used the excitation and ionization
      equilibrium  of Fe, Cr and Ti to derive $T_{\rm eff}$, log $g$  and $\xi$.

  HD 10132  has a significantly large radial velocity of $-70$ km s$^{-1}$. 
  Carbon is deficient, possibly caused by CN processing.
 Among $\alpha$ elements, sulfur shows a mild  enrichment relative to iron.
   Fe-peak elements and (within the measurement errors) s-process elements show
   near solar abundances. Similarly,
     HD 159251 shows near solar abundance for most of the elements.
     However, these abundances  are the only spectroscopic estimates made 
     based on a very  large number of lines. 
     These estimates 
    could be used for the calibration of photometric indices of cool stars. 
        
    Because of its binarity, HD 12533 was included in the barium star survey 
           of Za\v{c}s (1994), but it was not found to show significant 
            enhancement of Ba and other s-process elements. 
           Our present analysis uses better resolution spectra and 
            much larger spectral coverage,  and therefore includes 
           much larger number of lines for each species.   
           In spite of the low temperature, the spectrum is  not overcrowded. 
          We confirm a lack of enhancement for Ba and other s-process elements reported
          by Za\v{c}s and  find  
           near solar abundances for the other elements. We do not 
           confirm the [Fe/H]= $-0.4$ estimated by Za\v{c}s. Our derived temperature
         (4250~K) is slightly smaller than that derived by Za\v{c}s (4350~K)
           and we derive higher gravity (log $g$= 2.0) than that of  Za\v{c}s 
      (log $g$= 0.5).  In the light of the better quality of data used and the extensive 
       spectral coverage, the abundances presented in Table 4 are 
likely more accurate.

      \section {Conclusions}\label {conclusions} 

      The present work is a continuation of Paper I extending to          
  high temperature regimes of the post-AGB candidates.
    HD 27381 appears to be very similar to the post-AGB star HD 107369. 
    These two stars are candidates for a search for light variability
 and better spectral coverage as they appear intermediate between
    cooler post-AGB stars with exotic abundances and hot post-AGB stars.

 Another important finding of the present work is the abundance peculiarities   
   exhibited by HD 137569.  It shows selective depletion of the refractory 
  elements as seen in the case of HR 4049 and other well-known post-AGB stars and 
   many RV Tau stars. It is relatively cooler than the hot post-AGB stars studied 
   by Conlon et al. (1991) and McCausland et al. (1992) but two objects, 
   LS IV$-4^{o}.01$ and  LB 3193 which are of similar temperature,
  show a similar deficiency of Mg and Si. A large number of post-AGB stars
  showing depletion of refractory elements are found to be binaries.
  HD 137569 is known to be a spectroscopic binary but the lack of IR 
  fluxes of this object might be due to mass ejection having
  occurred long ago. The carbon deficiency observed for this star is
  not as large as that seen for the hot post-AGB stars. On the other hand 
  the derived C,N,O abundances are similar to PNe abundances. 
  HD 172324 is a moderately metal-poor, high velocity star that does not
  resemble post-AGB stars in abundance pattern, but
  deserves  continuous photometric and radial velocity monitoring 
  in search of binarity and/or pulsation.
  HD 10285 and HD 25291 do not appear to be post-AGB stars but show evidence of red giant mixing that has brought the products of the NeNa cycle to the surface. 
            The rest of the stars appear to be normal stars of 
           near solar composition.   

    Investigation of  high galactic latitude supergiants 
    with light variability could be quite rewarding even if the criteria 
  of two peaks in the SED  is not met. These might be post-AGB stars 
   with progenitors of lower masses. Here the transition time being larger, 
   these  stars might have 
  lost most of their envelope before moving towards the high temperature regime 
   and may be left with sparse and cold circumstellar matter (if present). 
  To the best of our knowledge BD$+$39$^{o}~$4926 is the only outlyer object
   among post-AGB stars, showing depletion of refractory elements but
  without IR excess. HD 137569 appears to be a new addition to the family.

      \acknowledgements 
      We express our gratitude to the anonymous referee for many important 
      suggestions. 
      We are indebted to Drs. E. Reddy and D. Young for obtaining some spectra for us. 
      We thank Dr. Gajendra Pandey for his help with high temperature low gravity 
       model atmospheres.
      We also acknowledge the support from DGAPA-UNAM grant through project IN110102 and are 
       thankful to the CONACyT (Mexico) and the Department of Science and Technology 
       (India), for the travel support and local hospitality respectively under 
       Indo-Mexican collaborative project 
       DST/INT/MEXICO/RP001/2001. 
      This work has made use of the SIMBAD database. 
      %

      \end{document}